\documentclass[
    ,final            
  ]
  {aipproc}

\layoutstyle{8x11single}

\usepackage{bm}

\newcommand{\dfrac}{\displaystyle\frac}

\begin{document}

\title{Exact conservation and breaking of pseudospin symmetry in single particle resonant states}


\author{Bing-Nan Lu}{
  address={State Key Laboratory of Theoretical Physics,
           Institute of Theoretical Physics, Chinese Academy of Sciences, 
           Beijing 100190, China}
}

\author{En-Guang Zhao}{
  address={State Key Laboratory of Theoretical Physics,
           Institute of Theoretical Physics, Chinese Academy of Sciences, 
           Beijing 100190, China}
  ,altaddress={Center of Theoretical Nuclear Physics, National Laboratory
           of Heavy Ion Accelerator, Lanzhou 730000, China}
}

\author{Shan-Gui Zhou}{
  address={State Key Laboratory of Theoretical Physics,
           Institute of Theoretical Physics, Chinese Academy of Sciences, 
           Beijing 100190, China}
  ,altaddress={Center of Theoretical Nuclear Physics, National Laboratory
           of Heavy Ion Accelerator, Lanzhou 730000, China}
}

\begin{abstract}
In this contribution we present some results on the study of pseudospin 
symmetry (PSS) in single particle resonant states. 
The PSS is a relativistic dynamical symmetry 
connected with the small component of the nucleon Dirac wave function. 
Many efforts have been made to study this symmetry in bound states.
We recently gave a rigorous justification of the PSS in single particle 
resonant states by examining the zeros of Jost functions corresponding 
to the small components of the radial Dirac wave functions and 
phase shifts of continuum states~\cite{Lu2012_PRL109-072501}.
We have shown that the PSS in single particle resonant states 
in nuclei is conserved when the attractive scalar and repulsive vector 
potentials have the same magnitude but opposite sign.
Examples of exact conservation and breaking of this symmetry 
in single particle resonances are given for spherical square-well 
and Woods-Saxon potentials.
\end{abstract}

\maketitle


\section{Introduction}

Since the concept of pseudospin (PS) was introduced 
more than 40 years ago~\cite{Arima1969_PLB30-517, Hecht1969_NPA137-129}, 
the pseudospin symmetry (PSS) in single particle states 
has been an interesting topic in nuclear physics~\cite{Dudek1987_PRL59-1405, 
Bahri1992_PRL68-2133, Blokhin1995_PRL74-4149}.
In 1997 it was found that the PSS in nuclei is a relativistic
symmetry which is exactly conserved when the scalar and
vector potentials have the same size but opposite sign, i.e.,
$\Sigma(r) \equiv S(r) + V(r) = 0$~\cite{Ginocchio1997_PRL78-436}.
Later it was shown that the PSS is exact under a less strict condition,
$d\Sigma(r)/dr = 0$~\cite{Meng1998_PRC58-R628, Meng1999_PRC59-154}.

By extending the study of the PSS to negative energy states, it was
found that the above mentioned conditions also result in the spin symmetry (SS) 
in anti-nucleon spectra which is much better developed than the PSS in nuclear 
single particle spectra~\cite{Zhou2003_PRL91-262501, He2006_EPJA28-265}.

The SS and PSS have been investigated extensively within the relativistic 
framework including the PSS in deformed 
nuclei~\cite{Lalazissis1998_PRC58-R45, Ginocchio2004_PRC69-034303, Setare2010_APPB41-2459},
the SS for $\bar\Lambda$ spectra in hyper-nuclei~\cite{Song2009_CPL26-122102, 
Song2010_ChinPhysC34-1425, Song2011_CPL28-092101},
and these symmetries in various potentials~\cite{Zhang2008_PRA78-040101, 
Zhang2009_PS80-065018, 
Akcay2009_IJMPC20-930, 
Ke2010_IJMPE25-1123, 
Wei2010_PLB686-288, Wei2010_EPJA43-185, 
Zhang2011_JMP52-053518, 
Aydogdu2011_PLB703-379, 
Chen2011_ChinPhysB20-062101,
Ikhdair2012_FBS53-473, 
Hamzavi2012_IJMPE21-1250097, 
Candemir2012_IJMPE21-1250060, 
Aydogdu2013_ChinPhysB22-010302}.
The readers are referred to Ref.~\cite{Ginocchio2005_PR414-165} for a review
and Refs.~\cite{Typel2008_NPA806-156, Leviatan2009_PRL103-042502, 
Long2010_PRC81-031302R, Lisboa2010_PRC81-064324, Li2011_ChinPhysC35-825, 
Liang2010_EPJA44-119,
Liang2011_PRC83-041301, Guo2012_PRC85-021302R, 
Castro2012_PRC86-052201, Castro2012_PRA86-032122, 
Liang2013_PRC87-014334} for recent progresses.
In particular, a very extensive overview was given in Ref.~\cite{Liang2013_PRC87-014334}
for the study of PSS and SS.

Quite recently, the node structure of radial Dirac wavefunctions in central 
confining potentials was studied and it was shown in a general way that 
it is possible to have positive energy bound solutions for these potentials
under the condition of exact PSS~\cite{Alberto2013_arXiv1302.2067}.
Note that there have been some investigations of PSS connected 
with some specific forms of confining 
potentials~\cite{Chen2003_CPL20-358, Chen2003_HEPNP27-324}.

In recent years, the study of nuclear single-particle resonant states, 
especially those in exotic nuclei with unusual $N/Z$ ratios
becomes more and more important~\cite{Yang2001_CPL18-196, 
Zhang2003_SciChinaG46-632,
Zhang2004_PRC70-034308, 
Zhang2008_PRC77-014312, Zhou2009_JPB42-245001, Fedorov2009_FBS45-191,  
Guo2010_PRC82-034318, Zhou2010_PRC82-011301R, Pei2011_PRC84-024311}. 
There have been some numerical studies on the PSS~\cite{Guo2005_PRC72-054319, Guo2006_PRC74-024320, 
Zhang2006_HEPNP30S2-97, Zhang2007_CPL24-1199} and
SS~\cite{Xu2012_IJMPEE21-1250096} in single particle resonant states.
In a recent work, we have shown that the PSS in single particle resonant states 
is also exactly conserved under the same condition for the PSS in bound states, 
i.e., $\Sigma(r) = 0$ or $d\Sigma(r)/dr = 0$~\cite{Lu2012_PRL109-072501}.  
In this contribution we will present some results given in Ref.~\cite{Lu2012_PRL109-072501}. 

\section{Justification of pseudospin symmetry in resonant states}


We gave a rigorous justification of the PSS in single particle 
resonant states by examining the asymptotic behaviour of 
the radial Dirac wave functions in Ref.~\cite{Lu2012_PRL109-072501};
more details will be given in a future publication~\cite{Lu2013_in-prep}.
Next we briefly mention the justification of the PSS 
in single particle resonant states under the condition 
$\Sigma(r) = 0$ or $d\Sigma(r)/dr = 0$.

In relativistic mean field models, the Dirac equation for a nucleon 
reads~\cite{Ring1996_PPNP37-193, Vretenar2005_PR409-101, Meng2006_PPNP57-470}
\begin{equation}
 \left[ \bm{\alpha} \cdot \bm{p}
      + \beta \left( M + S(\bm{r}) \right)
      + V(\bm{r})
 \right] \psi(\bm{r})
 = \epsilon \psi(\bm{r})
 ,
~\label{Eq:diraceq}
\end{equation}
where $\bm{\alpha}$ and $\beta$
are the Dirac matrices and $M$ is the nucleon mass.
For a spherical nucleus, the Dirac spinor 
\begin{equation}
 \psi (\bm{r}) = \frac{1}{r}
  \left(
   \begin{array}{c}
    i F_{n\kappa}(r)         Y_{jm}^{l}(\theta,\phi )        \\
    - G_{\tilde{n}\kappa}(r) Y_{jm}^{\tilde{l}}(\theta,\phi )
   \end{array}
  \right) ,
 \label{eq:SRHspinor}
\end{equation}
where $Y_{jm}^{l}(\theta,\phi)$ is the spin spherical harmonic. 
$F_{n\kappa }(r)/r$ and $G_{\tilde{n}\kappa}(r)/r$ are the radial 
wave functions for the upper and lower components with $n$ and $\tilde{n}$ radial nodes.
$\kappa = (-1)^{j+l+1/2}(j+1/2)$
and $\tilde{l}=l-{\mathrm{sign}}(\kappa)$.
For brevity we will omit the subscripts from $F(r)$ and $G(r)$ whenever no confusion arises.
The radial Dirac equation is a first order coupled equation 
and can be rewritten as two decoupled second order differential ones. 
For the small component to which the PSS is directly connected, 
the second order differential equation reads,
\begin{eqnarray}
 \left[
   \frac{d^{2}}{dr^{2}}
  -\frac{1}{M_{-}(r)}\dfrac{d\Sigma(r)}{dr}\frac{d}{dr}
  -\dfrac{\tilde{l}(\tilde{l}+1)}{r^{2}}
  + \dfrac{1}{M_{-}(r)}
    \frac{\kappa}{r}\dfrac{d\Sigma(r)}{dr}
  -  M_+(r) M_-(r)
 \right] G(r)
 & = & 
 0 , 
 \label{eq:G}
\end{eqnarray}
where $M_{+}(r) \equiv M + \epsilon - \Delta(r)$ and $M_{-}(r) \equiv M - \epsilon + \Sigma(r)$.

For the continuum in the Fermi sea, i.e., $\epsilon \geq M$, there exist two 
independent solutions for Eq.~(\ref{eq:G}). 
The physically acceptable solution is the one that vanishes at the origin
and behaves like 
$j_{\tilde{l}}(pr)$ as $r \rightarrow 0$~\cite{Taylor1972},
\begin{equation}
  \lim_{r \rightarrow 0} G(r)/j_{\tilde{l}}(pr) = 1,
  \ p = \sqrt{\epsilon^2 - M^2} .
 \label{eq:regularsolution}
\end{equation}

The resonance parameters are connected with the asymptotic behavior of 
the regular solution as $r\rightarrow\infty$. 
At large $r$ the potentials for neutrons vanish and the wave functions oscillate, 
Eq.~(\ref{eq:G}) becomes a Ricatti-Bessel equation with angular momentum $\tilde{l}$ and
the solution can be written as a combination of the
Ricatti-Hankel functions,
\begin{equation}
 G(r) 
 = \frac{i}{2} \left[
                       \mathcal{J}_{\kappa}^{G}(p)     h_{\tilde{l}}^{-}(pr) 
                     - \mathcal{J}_{\kappa}^{G}(p)^{*} h_{\tilde{l}}^{+}(pr)
               \right],
   \ r\rightarrow\infty
 ,
\end{equation}
where $\mathcal{J}_{\kappa}^{G}(p)$ is the Jost function for the
small component and $h_{\tilde{l}}^{\pm}(pr)$  the Ricatti-Hankel functions.
It has been argued that in nuclei the Jost function is an analytic function of $p$ 
and can be analytically continued to a large area in the complex $p$ 
plane~\cite{Lu2012_PRL109-072501}. 
The zeros of the Jost function $\mathcal{J}_{\kappa}^{G}(p)$ 
on the complex momentum plane correspond to bound states or resonances:
The zeros on the positive imaginary axis of the $p$ plane represent 
bound states of the original eigenvalue problem, 
while those on the lower $p$ plane and 
near the real axis correspond to resonant states. 
The resonance energy $E_{{\rm res}}$ and width $\Gamma_{{\rm res}}$
are determined by the relation $E=E_{{\rm res}}-i{\Gamma_{{\rm res}}}/{2}=\sqrt{p^{2}+M^{2}}$. 
By examining the zeros of the Jost function we studied the bound
and resonant states on the same footing~\cite{Lu2012_PRL109-072501}.

In the PSS limit, Eq.~(\ref{eq:G}) is reduced as
\begin{equation}
 \left[
          \frac{d^{2}}{dr^{2}}
        - \frac{\tilde{l}(\tilde{l}+1)}{r^{2}}
        + \left(
                \epsilon-M
          \right)
          M_+(r)
 \right]  G(r)
 =  0
 .
~\label{eq:PSLeqG}
\end{equation}
For the continuum $\epsilon$ can be any value $\ge M$ and 
one can mainly focus on wave functions and their asymptotic behavior.
For PS doublets with different quantum numbers $\kappa$ and $\kappa'$ with 
$\kappa^{\prime}=-\kappa+1$, the small components satisfy the same equation because 
they have the same pseudo-orbital angular momentum $\tilde{l}$~\cite{Ginocchio1997_PRL78-436}.
In particular, for continuum states, $G_{\kappa}(\epsilon,r)=
G_{\kappa^{\prime}}(\epsilon,r)$ for any energy $\epsilon$. 
Because the definition of the Jost function $\mathcal{J}_{\kappa}^{G}(p)$ only
depends on the asymptotic behavior of the small component, we have
$\mathcal{J}_{\kappa^{\prime}}^{G}(p)=\mathcal{J}_{\kappa}^{G}(p)$
on the positive real axis. 
This equivalence can be generalized into the complex $p$ plane 
due to the uniqueness of the analytic continuation. 
Thus the zeros are the same for $\mathcal{J}_{\kappa^{\prime}}^{G}(p)$ and 
$\mathcal{J}_{\kappa}^{G}(p)$: If there exists a resonant state with 
energy $E_{{\rm res}}$ and width $\Gamma_{{\rm res}}$ and the quantum number $\kappa$, 
there must be another one with the same energy and width and quantum number $\kappa^{\prime}$. 
That is to say, the PSS in single particle resonant states 
in nuclei is exactly conserved when the attractive scalar and repulsive vector 
potentials have the same magnitude but opposite sign~\cite{Lu2012_PRL109-072501}.

In scattering theories, one can also determine resonance parameters
from the change of cross section or phase shift which 
give us more insights into the resonant phenomena. 
In Ref.~\cite{Lu2012_PRL109-072501}, the PSS in resonant states has
also been studied by examining the phase shift. 
For simplicity, we do not give the details here. 

\section{Numerical test of pseudospin symmetry in resonant states}

For PS doublets of single particle resonant states in nuclei, not only
the energies, but also the widths are exactly the same in the PSS limit.
Similar to what happens in bound states, when the PSS limit is 
not realized, the PSS in resonant states is broken.
In Ref.~\cite{Lu2012_PRL109-072501} we used a solvable model
and the Woods-Saxon potentials
to illustrate the conservation and the breaking of the PSS in resonant states. 

We considered that $\Sigma(r)$ and $\Delta(r)$
are both spherical square-well potentials,
\begin{equation}
 \Sigma(r)  =  \left\{ \begin{array}{c}
                                C,\qquad r<R ,\\
                                0,\qquad r\geq R ,\end{array}
                 \right.  
 \ \ \
 \Delta(r)  =  \left\{ \begin{array}{c}
                                D,\qquad r<R ,\\
                                0,\qquad r\geq R ,\end{array}
                 \right.
 \label{eq:square_well}
\end{equation}
where $C$ and $D$ are constants and $R$ is the width.
For $r\neq R$, Eq.~(\ref{eq:G}) reads,
\begin{equation}
 \left[
           \frac{d^{2}}{dr^{2}}
        -  \frac{\tilde{l}(\tilde{l}+1)}{r^{2}}
        - M_+(r) M_-(r)
          \right] G(r)
   =      0
 .
\end{equation}
The regular solution of this equation is just a combination of the Ricatti-Bessel functions,
\begin{eqnarray}
 G(r) & = & (p/k)^{\tilde{l}+1} j_{\tilde{l}} \left( kr \right),\ r<R,
~\label{Eq:Gr-} 
 \\
 G(r) & = &   \frac {i}{2} \left[
                 \mathcal{J}_{\kappa}^{G}(p)     h_{\tilde{l}}^{-}(pr)
               - \mathcal{J}_{\kappa}^{G}(p)^{*} h_{\tilde{l}}^{+}(pr)
                            \right],
 r\geq R, 
~\label{Eq:Gr+} 
\end{eqnarray}
with $k = \sqrt{\left( \epsilon - C - M \right) \left( \epsilon - D + M \right)}$.
The coefficient $(p/k)^{l+1}$ is inserted in accordance with Eq.~(\ref{eq:regularsolution})
and $\hat{j}_l(z) \propto z^{l+1}$ as $z \rightarrow 0$.

The Jost function for the small component is derived as~\cite{Lu2012_PRL109-072501},
\begin{eqnarray}
 \mathcal{J}_{\kappa}^{G}(p)  
 & = &  
 -\frac{p^{\tilde{l}}}{2ik^{\tilde{l}+1}}
  \left[
        j_{\tilde{l}}(kR) p h_{\tilde{l}}^{+\prime}(pR)
          -  kj_{\tilde{l}}^{\prime}(kR) h_{\tilde{l}}^{+}(pR)
  \right.
  \left.
          -  \frac{C}{\epsilon-M-C} \left(
                                           kj_{\tilde{l}}^{\prime}(kR)
                                         - \frac{\kappa}{R} j_{\tilde{l}}(kR)
                                    \right) h_{\tilde{l}}^{+}(pR)
  \right]
 .
~\label{Eq:JostG}
\end{eqnarray}
The Jost functions $\mathcal{J}_{\kappa}^{G}(p)$ and
$\mathcal{J}_{\kappa^{\prime}}^{G}(p)$ differ
only in the part containing $C$ because they have the same $\tilde{l}$.
Therefore, in the PSS limit, i.e., $C=0$,  
$\mathcal{J}_{\kappa}^{G}(p)=\mathcal{J}_{\kappa^{\prime}}^{G}(p)$.
Consequently the PSS is conserved both in bound states and in resonant states.
If $C\ne 0$, the PSS is broken and we can study the PS splitting 
of the energy and the width for resonant states.
Due to the special form of the spherical square-well potentials,
the PSS-breaking term is separated from the PSS-conserving term in the Jost function, 
which makes the study of the conservation or the breaking of the PSS very convenient.

\begin{figure}
\begin{centering}
\includegraphics[width=0.6\columnwidth]{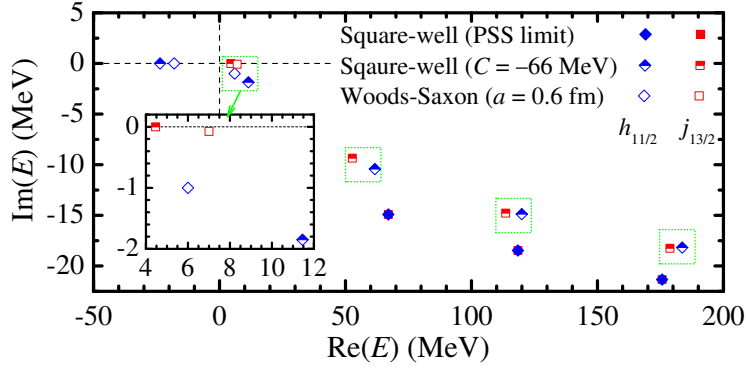}
\par\end{centering}
\caption{(Color online)
The zeros of the Jost function $\mathcal{J}_{\kappa}^{G}$ on the complex energy plane
in square-well potentials (\ref{eq:square_well}) with 
$C=0$ (solid symbols) and $C=-66$ MeV (half-filled symbols) 
for PS partners $h_{11/2}$ (diamond) and $j_{13/2}$ (square). 
The results with Woods-Saxon-like scalar and vector potentials are also
shown as open symbols.
Take from Ref.~\protect\cite{Lu2012_PRL109-072501}.
}
~\label{fig:zeros}
\end{figure}

Figure~\ref{fig:zeros} show solutions in the complex energy plane 
for PSS doublets with $\tilde{l}=6$, i.e., $h_{11/2}$ with $\kappa = -6$ 
and $j_{13/2}$ with $\kappa' = 7$ for square-well potentials 
with $D=650$ MeV and $R=$ 7 fm~\cite{Lu2012_PRL109-072501}.
In the PSS limit, i.e., $C=0$, all the roots locate in the lower 
half plane and there are no bound states. 
The conservation of the PSS for single particle resonant states is clearly seen.
When $C=-66$ MeV, there is one bound state only for $h_{11/2}$. 
Three pairs of PS partners of resonant states are shown by
half-filled diamonds and squares.
One finds the breaking of the PSS both in the bound states and in the resonant states.
For PS doublets with other values of $\tilde{l}$, we observed similar 
behaviors concerning the exact conservation and the breaking of the PSS.
We also studied resonances in Woods-Saxon-like potentials, 
$W(r) = W_0 / (1+\exp[(r-R)/a])$ ($W=V$ or $S$)
with parameters connected with $^{208}$Pb given in Ref.~\cite{Guo2005_PRC72-054319}: 
the depths $V_0 - S_0$ = 650 MeV and $V_0 + S_0$ = $-66$ MeV, the diffusivity parameter 
$a=0.6$ fm, and $R=7$ fm. 
Resonance parameters are obtained with the real stabilization 
method~\cite{Zhang2008_PRC77-014312, Zhou2009_JPB42-245001}. 
The results are shown as open diamonds and squares for $h_{11/2}$ and $j_{13/2}$, respectively.
It is found that splittings of energy and width both become smaller
compared with the results with the square-well potentials.
The reason is that the derivative of $\Sigma(r)$ is smaller due to 
a non-zero diffusivity parameter.

\section{Summary and perspectives}

We showed that the PSS in single particle
bound and resonant states in nuclei can be investigated on the 
same footing within the relativistic framework
by examining the zeros of Jost functions corresponding to
small components of nucleon Dirac wave functions.
In the PSS limit, i.e.,
the attractive scalar and repulsive vector 
potentials have the same magnitude but opposite sign,
small components of PS doublets are exactly the same. 
Thus Jost functions describing the asymptotic behavior
of the radial wave functions are identical to each other. 
When analytically continued to complex momentum plane, 
the resonant states, showing themselves as zeros of 
the Jost functions, are always paired in the PSS limit,
which leads to the exact PSS. 
The conservation of the PSS in the PSS limit
was also justified by examining the phase shift of continuum states. 
When leaving the PSS limit, the PSS in resonant states is broken.
These conclusions were tested for single particle resonances 
in spherical square-well and Woods-Saxon potentials.

Finally we would like to mention that the work presented
in Ref.~\cite{Lu2012_PRL109-072501} and here extends 
the study of relativistic symmetries to resonant states.
There are many related open questions: 
\begin{itemize}
\item
Are there any experimental evidences of the PSS or SS in single particle resonant states?
\item
Having in mind that the centrifugal barriers are quite different
for PS doublets of single particle resonant states, 
how to understand intuitively that their widths are exactly the same in the PSS limit?
\item 
Is the breaking mechanism of the PSS or SS in single particle resonant states
the same as that in bound states?
\item
What about the relations between the wave functions of resonant doublet states?
\item
How about the PSS or SS in resonant states in anti-nucleon or anti-hyperon spectra?
\item
How about the PSS or SS in resonant states in deformed systems?
\item
What is the effect of the Coulomb interaction, for example, for protons, on the PSS
or SS symmetries?
\item
How about the case that one of the doublet states is bound and the other is in continuum?
\item
$\cdots$
\end{itemize}
We are working on some of these problems~\cite{Lu2013_in-prep} and many others 
are awaiting much further exploration.


\begin{theacknowledgments}
This work has been supported by 
Major State Basic Research Development Program of China (Grant No. 2013CB834400), 
National Natural Science Foundation of China (Grant Nos. 11121403, 11175252, 
11120101005, 11211120152, and 11275248),
Knowledge Innovation Project of Chinese Academy of Sciences (Grant
Nos. KJCX2-EW-N01 and KJCX2-YW-N32). 
The results described in this paper are obtained on the ScGrid of
Supercomputing Center, Computer Network Information Center of Chinese Academy
of Sciences.
\end{theacknowledgments}



\bibliographystyle{aipproc}   


\end{document}